\documentclass[twocolumn]{aa}
\usepackage[usenames]{color}
\usepackage{mathrsfs}
\usepackage{graphicx}
\usepackage{natbib}
\usepackage{enumitem}
\usepackage{hyperref}
\usepackage[normalem]{ulem}
\usepackage{amsmath}

\newcommand{\grav}{{{g_0}}}

\newcommand{\Mm}{{\mathrm{\, Mm}}}
\newcommand{\kms}{{\mathrm{\, km \,s^{-1}}}}

\newcommand{\rsun}{R_\odot}
\newcommand{\psun}{P_\odot}
\newcommand{\pcutoff}{P_\mathrm{cutoff}}
\newcommand{\pslow}{P_\mathrm{slow}}
\newcommand{\newrlim}{\widehat{R}_\mathrm{lim}}
\newcommand{\gvalue}{274~\mathrm{m\, s^{-2}}}

\DeclareGraphicsExtensions{.pdf,.png,.jpg}

\newcommand{\mins}{{\mathrm{\, minutes}}}

\begin{document}
\title{Extension and validation of the pendulum model for longitudinal solar prominence oscillations}
\author{M. Luna, J. Terradas, J. Karpen \& J. L. Ballester}

\date{Received <date> /
Accepted <date>}

\author{M. Luna \inst{1,2}
\and J. Terradas \inst{1,2}
\and J. Karpen \inst{3}
\and J. L. Ballester \inst{1,2}
     }

  \institute{Departament F{\'i}sica, Universitat de les Illes Balears, E-07122 Palma de Mallorca, Spain
       \email{manuel.luna@uib.es}
       \and
       Institute of Applied Computing \& Community Code (IAC$^3$), UIB, Spain
       \and
       Heliophysics Science Division, NASA Goddard Space Flight Center, 8800 Greenbelt Road, Greenbelt, MD 20771, USA
       }

\titlerunning{Extension and validation of the pendulum model}
\authorrunning{Luna, Terradas, Karpen and Ballester}

\abstract {Longitudinal oscillations in prominences are common phenomena on the Sun. These oscillations can be used to infer the geometry and intensity of the filament magnetic field. Previous theoretical studies of longitudinal oscillations made two simplifying assumptions: uniform gravity and semi-circular dips on the supporting flux tubes. However, the gravity is not uniform and realistic dips are not semi-circular.}
{To understand the effects of including the nonuniform solar gravity on longitudinal oscillations, and explore the validity of the pendulum model with different flux-tube geometries.}
{We first derive the equation describing the motion of the plasma along the flux tube including the effects of nonuniform gravity, yielding corrections to the original pendulum model. We also compute the full numerical solutions for the normal modes, and compare them with the new pendulum approximation.}
{We have found that the nonuniform gravity introduces a significant modification in the pendulum model. We have also found a cut-off period, i.e. the longitudinal oscillations cannot have a period longer than 167 minutes. In addition, considering different tube geometries, the period depends almost exclusively on the radius of curvature at the bottom of the dip.}
{We conclude that nonuniform gravity significantly modifies the pendulum model. These corrections are important for prominence seismology, because the inferred values of the radius of curvature and minimum magnetic-field strength differ substantially from those of the old model. However, we find that the corrected pendulum model is quite robust and is still valid for non-circular dips.}


\maketitle

\section{Introduction}\label{sec:intro}

Solar prominences are very dynamical objects with many kinds of motions, including oscillations and wave phenomena. A common type of motion is large-amplitude oscillations, in which a large portion of the filament oscillates coherently with velocities above $10\kms$.
These oscillations have been reported since the first half of the twentieth century \citep[see the review of][]{arregui_prominence_2018}. 
\citet{jing_periodic_2003} identified large-amplitude \emph{longitudinal} oscillations (LALOs) along prominence threads after a nearby energetic event and many more LALOs have been reported since then
\citep[see][]{luna_gong_2018}. LALOs are excited by eruptive flares near the filament \citep[e.g.,][]{jing_periodic_2003,vrsnak_large_2007,zhang_observations_2012}, jets \citep{luna_observations_2014,zhang_simultaneous_2017}, or coronal shocks \citep{shen_simultaneous_2014}. However, in many cases the trigger agent has not been identified. LALO periods range from a few tens of minutes up to 160 minutes \citep{jing_periodic_2006}, but most of the reported events have periods around one hour \citep{luna_gong_2018}.
LALOs are important because they can be used to infer the geometry and intensity of the filament magnetic field  \citep{luna_large-amplitude_2012,zhang_observations_2012,luna_observations_2014,luna_large-amplitude_2017}. Prominence magnetic structure is a long-standing question, because we can only observe the photospheric field directly and competing models have been proposed. Because prominences reside in filament channels, which are the sources of all major solar eruptions, understanding the full coronal structure of these channels is crucial for resolving the fundamental physical mechanism(s) responsible for eruptions and resulting space weather. One key consequence of the LALO phenomenon is that the flux tubes supporting and guiding the prominence plasma must contain dips: concave upward regions in which Lorentz forces balance the gravitational force. This constraint led to a surprisingly accurate ``pendulum'' model for LALO motions, and a new prominence-seismology technique for deriving the dip radius of curvature and the minimum magnetic-field strength in the dip \citep{luna_large-amplitude_2012,luna_observations_2014}.

Previous theoretical studies of longitudinal oscillations made two simplifying assumptions that we explore further in this paper: uniform gravity and semi-circular dips on the supporting flux tubes. The first assumption is to consider gravity as a uniform vector with a value of $\gvalue$, which is vertical at all points of the studied volume. This approximation is widely used in theoretical studies when the region studied is a small portion of the Sun.
Examples of theoretical studies of longitudinal oscillations using uniform gravity include \citet{luna_large-amplitude_2012,zhang_parametric_2013,%
luna_robustness_2016,terradas_solar_2016}, and \citet{liakh_numerical_2020}.
However, gravity is not uniform and depends on position. The gravity vector always points to the centre of the Sun and its intensity decreases with distance. 
As we will demonstrate in this paper, the spatial dependence of gravity can strongly influence longitudinal oscillations of solar prominences.

Second, previous models assumed semi-circular dips of the field lines supporting the prominence threads \citep{luna_large-amplitude_2012,luna_effects_2012,luna_effects_2016}. In this geometry, the radius of curvature along the dip, $R_0$, is constant. However, in a realistic situation, the curvature is not likely to be constant. The radius of curvature depends on the position along the dip $R=R(s)$ with $R_0=R(s=0)$, where $s$ is the coordinate along the flux tube and $s=0$ at the bottom of the dip. In this situation, the relation between the period and the radius of curvature is not immediately obvious. Numerical simulations of longitudinal oscillations in different flux-tube geometries \citep[see e.g.,][]{zhang_parametric_2013,luna_robustness_2016,liakh_large-amplitude_2021} have shown that the pendulum model is still valid considering the radius of curvature at the bottom of the dip, $R_0$, or a curvature averaged around the centre of the dip. In the present study, we investigate the robustness of the model and identify under which conditions the pendulum model remains valid.

In this study, we consider that LALOs can be described by linear magnetoacoustic-gravitational modes  \citep{luna_effects_2012,  terradas_magnetohydrodynamic_2013}. This allows us to study LALOs analytically and find useful relationships that apply to prominence seismology. However, the displacements and velocities in LALOs are large and nonlinear effects may be relevant. Nonlinear effects will be considered in a future study.
In this paper, we present the effects of including the nonuniform solar gravity and then explore the validity of the pendulum model with different flux-tube geometries. In \S\ref{sec:model} we introduce the model that describes the equilibrium and dynamics of a prominence thread with the nonuniform gravity incorporated. In \S\ref{sec:circulargeometry}, the longitudinal oscillations in a flux tube with semi-circular dip geometry are studied to find the influence of curvature of the solar surface. In \S\ref{sec:tube-geometries}, the study is extended to alternative flux-tube geometries to investigate the validity of the pendulum model in these geometries. In \S\ref{sec:seismology} we discuss the influence of the solar-surface curvature on the determination of the radius of curvature of the dips and the minimum magnetic-field strength of the prominences using seismology. Finally, in \S\ref{sec:conclusions} the conclusions of this investigation are summarised.

\section{The model}\label{sec:model}

Here we derive the equation that describes the motion of the plasma velocity perturbations including the intrinsic spatial variations of the solar gravity. We also show the different flux-tube geometries used to determine the influence of the dip shape on the oscillations. Finally we consider the equilibrium of the plasma along the flux tube. We assume that the central part is filled with cool prominence plasma representing a thread. The temperature increases sharply at its sides, reaching a coronal temperature that applies to the rest of the tube up to the footpoints.

\subsection{Non-uniform gravity considerations}\label{subsec:non-uniform-gravity-considerations}

The solar gravity is given by the expression
\begin{align}\label{eq:gravity}
\vec{g}=-\grav \, \left( \frac{\rsun}{\rsun+h} \right)^2 \, \vec{\hat{r}} \, , 
\end{align}
where $\grav=\gvalue$ is the acceleration at the solar surface, $h$ is the height with respect to the solar surface and $\vec{\hat{r}}$ is the radial unit vector pointing in the (minus) direction of gravity.
Most prominences are located in the low corona, in which $h \ll \rsun$. Thus Eq. \eqref{eq:gravity} can be approximated by
\begin{equation}\label{eq:gravity-approx}
\vec{g}\approx-g_{0} \, \vec{\hat{r}} \, .
\end{equation}
This indicates that, for filaments in the low corona, the main spatial variation of gravity is due to its changing direction, not its intensity. The novelty of this work is to introduce this intrinsic variation of the solar gravity with position. For the study of longitudinal oscillations we are interested into the projection of the gravity along the magnetic field of solar prominences, $g_\parallel$. It is important to note that, in the uniform gravity situation, $g_\parallel$ changes along the field line because the magnetic-field direction changes. However, in the nonuniform situation $g_\parallel$ also changes with the intrinsic changes of the direction of $\vec{g}$ given by Eq. \eqref{eq:gravity-approx}.

\subsection{The governing equation}
The oscillations along the magnetic field in 2D configurations were described by \citet{goossens_existence_1985} as so-called slow continuum modes, also denoted magnetoacoustic-gravity modes \citep{terradas_magnetohydrodynamic_2013}. In contrast to usual MHD slow modes, both gas-pressure gradients and gravity can be the restoring forces of these modes. 
We assume a low-$\beta$, adiabatic plasma confined in uniform cross-section flux tubes along a static magnetic field, with with no heating or radiation terms, similar to \citet{luna_effects_2012} (hereafter called Paper I). 
The novelty is that we consider the intrinsic spatial variations of the solar gravity field using Eq. \eqref{eq:gravity-approx}. With these assumptions, from the ideal MHD equations we recover our previous expression (Eq. (5) in Paper I),
\begin{equation}\label{eq:wave-eq}
\frac{\partial^{2} v}{\partial t^{2}} - c_{s}^{2} \frac{\partial^{2} v}{\partial s^{2}} =\gamma g_{\parallel} \frac{\partial v}{\partial s} + v \frac{\partial g_{\parallel}}{\partial s} ~,
\end{equation}
where $c_{s}^{2}=\gamma \, p/\rho$ is the sound speed and $p$ and $\rho$ are the equilibrium gas pressure and density respectively. 
Note that $g_{\parallel}$ depends on the position $s$ due to two contributions: the change in the projection of the gravity along the field line, and the intrinsic variation of the gravity field (Eq. \eqref{eq:gravity-approx}) that is  introduced in this work.
Assuming an harmonic time dependence, $e^{i\omega t}$, the governing equation becomes
\begin{equation}\label{eq:wave-eq-harmonic}
c_{s}^{2} \frac{\partial^{2} v}{\partial s^{2}} + \gamma g_{\parallel} \frac{\partial v}{\partial s} +\left(\omega^2 + \frac{\partial g_{\parallel}}{\partial s}\right) v =0 ~.
\end{equation}
Both terms $g_{\parallel}$ and $\partial g_{\parallel}/\partial s$ depend on the geometry of the flux tube. For a semi-circular dip it is still possible to gain some insight into the effect of the intrinsic spatial variation of gravity (see \S \ref{sec:circulargeometry}) using an analytical approach. However, it is difficult to find an analytic expression for both terms for more general geometries, so we solve Eq. \eqref{eq:wave-eq-harmonic} numerically for different shapes of the flux tubes.

\subsection{Flux tube models}\label{subsec:geometry}
\begin{figure}[!h]
	\centering\includegraphics[width=0.56\textwidth]{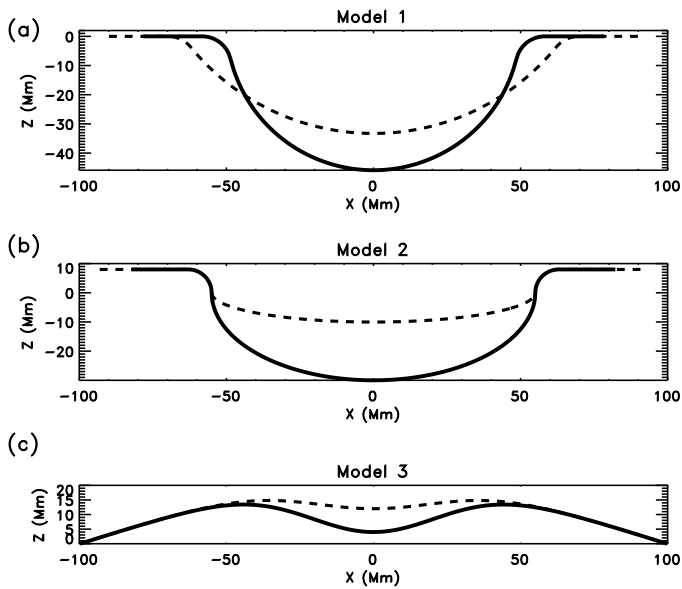}
	\caption{Flux tube geometries considered in this work: (a) semi-circular dip, (b) elliptical dip, and (c) sinusoidal. The solid line is the flux tube with the smallest radius of curvature. The dashed line is a case with a radius of curvature intermediate between the minimum and the maximum. In both Models 1 and 2 the end points of the flux-tube components change in order to keep the total length of the field line constant.}
\label{fig:fluxtube-geometries}
\end{figure}

In this work we explore the possible dependence of longitudinal oscillations on the geometry of the field lines. Although there are many possibilities, we consider three geometries for the flux tubes shown in Fig. \ref{fig:fluxtube-geometries}. We have chosen these cases because they are described by functions of a few parameters that allow us to control the shape of the flux tube. The first one (Model 1, Fig. \ref{fig:fluxtube-geometries}(a)) consists of a central semi-circular segment of radius $R$. On both sides, there is a straight part representing the non-dipped part of the tube. The semi-circular dip and the straight parts are joined smoothly with a small arc with $R_{small}=10\Mm$ that is tangent to the central and the straight parts. We have considered a range of $R$ from 42 to 1000 Mm. The lengths of the different parts of the piecewise flux tube are changed to keep the total half-length of the tube to $L_{1/2}=100\Mm$.

Model 2 consists of a central semi-elliptical segment with major axis $a=55\Mm$ parallel to $x$ (Fig. \ref{fig:fluxtube-geometries}(b)); minor axis, $b$, is parallel to $z$ and ranges from $b=3$ to 55. The tube also has two straight, horizontal segments as in Model 1. The dipped and straight parts are joined smoothly with an arc of $R_\mathrm{small}=8\Mm$. The length of the straight part is varied to keep the half-length of the tube to $L_{1/2}=100\Mm$. The radius of curvature of the dipped part depends on the parameter $b$. At the bottom of the dip (i.e. the central position), the radius of curvature is
\begin{equation}\label{eq:radiusofcurvature-ellipse}
R_0=\frac{a^2}{b} \ .
\end{equation}
With the parameters stated above for Model 2, $R_0$ ranges from 37 to 1000 Mm.

Model 3 is the most realistic because the footpoints are anchored in the solar surface. The flux tube consists of a combination of sinusoidal functions
\begin{equation}
z(x)=A_0 \cos\left(\frac{2 \pi x}{\lambda_0}\right)-A_1 \left[\cos\left(\frac{2 \pi x}{\lambda_1}\right)+1\right]^n \, ,
\end{equation}
where $\lambda_0=399\Mm$, $\lambda_1=\lambda_0/2$, $n=4$, $A_0=20\Mm$ and $A_1=0.18-1\Mm$. The $z$-position of the dip changes with $A_1$ as $z_\mathrm{dip}=A_0-A_1 \, 2^n$ and ranges from 4 to 17 Mm. With the selected parameters the half-length of the tube is more or less constant, $L_{1/2}=100\Mm$. The radius of curvature at the bottom of the dip is given by
\begin{equation}\label{eq:radiusofcurvature-sinusoidal}
	\frac{1}{R_0}=\frac{2^{n+1}n A_1 \pi^2}{\lambda_1^2}-\frac{4 A_0 \pi^2}{\lambda_0^2} \, .
\end{equation}
With the parameters considered in this work, Model 3 has $R_0=55 - 1000 \Mm$.

The plasma equilibrium is stratified following the equation
\begin{equation}
\frac{d p}{d s} = \rho g_{\parallel} \, .
\end{equation}
The projected gravity, $g_{\parallel}$, is given by Eq. \eqref{eq:gravity-approx} where the effect of the change of direction is incorporated. 
However, note that the governing Eq. \eqref{eq:wave-eq-harmonic} only depends on the sound speed $c_s$, which is proportional to the square root of the temperature. 
To solve Eq. \eqref{eq:wave-eq-harmonic}, we calculate $c_s$ from the following temperature profile:
\begin{equation}\label{eq:temperature}
{\scriptsize T(s)=
\begin{cases} 
 T_\mathrm{p}\, , & \text{if $|s| \leq -l_{1/2}$} \\  
T_{+}  +T_{-} \cos\left[ \frac{\pi}{l_\mathrm{tr}} (|s|-l_{1/2}) \right]   \, , & \text{if $l_{1/2}<|s|\le l_{1/2}+l_\mathrm{tr}$} \\  
T_\mathrm{c}   \, ,  & \text{if $|s|\le l_{1/2}+l_\mathrm{tr}$}  
 \end{cases}  }
\end{equation}
where $T_\mathrm{p}= 6\times10^3 K$, $T_\mathrm{c}=10^6 K$, $T_{+}=\left(T_\mathrm{c} + T_\mathrm{p} \right)/2$, and $T_{-}=\left(T_\mathrm{c} - T_\mathrm{p} \right)/2$. $T_\mathrm{p}$ and $T_\mathrm{c}$ are the prominence and corona equilibrium temperatures respectively. The assumed half-length of the prominence thread is $l_{1/2}=5\Mm$ and the temperature changes smoothly from prominence to coronal temperatures within a distance $l_\mathrm{tr}=1\Mm$ at both ends of the thread. We have tested different sizes for $l_\mathrm{tr}$ and its influence on the results is small.

\section{Semi-circular geometry}\label{sec:circulargeometry}

The semi-circular dip geometry is the simplest, and has been used in previous LALO research  \citep[e.g.][]{luna_effects_2012,ruderman_damping_2016}. Now we show how the intrinsic variations of the solar gravity in this geometry introduce corrections in the longitudinal oscillations.

\subsection{Pendulum approximation}
\label{subsec:pendulumapproximation}

\begin{figure}[!h]
	\centering\includegraphics[height=0.5\textwidth]{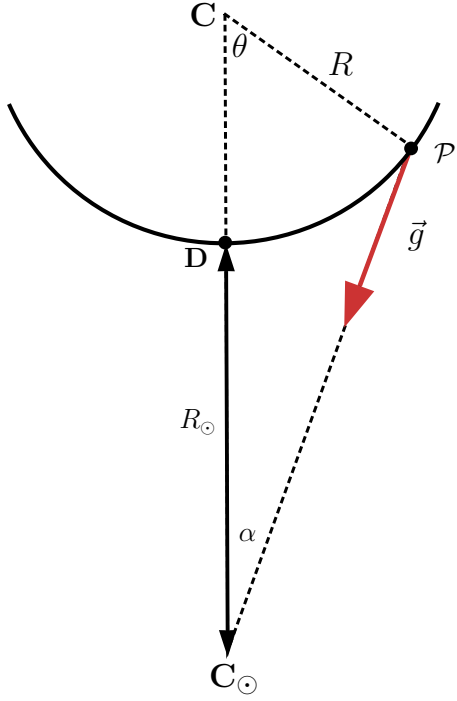}
	\caption{Sketch of a dip with circular shape (solid line) with radius $R$. The red arrow shows the solar gravity on a point of the tube ($\mathcal{P}$) pointing to the center of the Sun, $C_\odot$.\label{fig:sketch}}
\end{figure}

Fig. \ref{fig:sketch} shows the very simplified situation of a semi-circular dip of radius $R$ (thick solid curve). The center of curvature of the dip segment is $C$.

Many prominences are located in the low corona, so their heights above the surface, $h$, are small relative to $\rsun$. The distance from the Sun's centre, $C_\odot$ to the bottom of the dip, $D$, is $d\left(C_\odot,D\right)=\rsun+h$. However, $h/\rsun\ll1$ so $d\left(C_\odot,D\right)\approx\rsun$. This is equivalent to considering that the lower part of the dip is in contact with the surface as shown in the sketch (Fig. \ref{fig:sketch}). The gravity always points to the solar center, $C_\odot$. The relation between the angles $\theta$ and $\alpha$ is 
\begin{equation}\label{eq:relation-alpha-theta}
\alpha (\theta)= \arctan{\left(\frac{R \sin \theta}{R+R_{\odot}-R \cos \theta }\right)} \, .
\end{equation}
Any fluid element with position $\mathcal{P}$ moves along the semi-circular dip. We expect that the distance of any fluid element from the central position is small in comparison with the solar radius. Thus $R \sin \theta \ll R_{\odot}$. Therefore Eq. \eqref{eq:relation-alpha-theta} can be approximated as
\begin{equation}\label{eq:relation-alpha-theta-approximated}
\alpha (\theta)= \frac{R}{R_\odot} \theta \, . 
\end{equation}

\begin{figure}[!h]
\centering\includegraphics[width=0.45\textwidth]{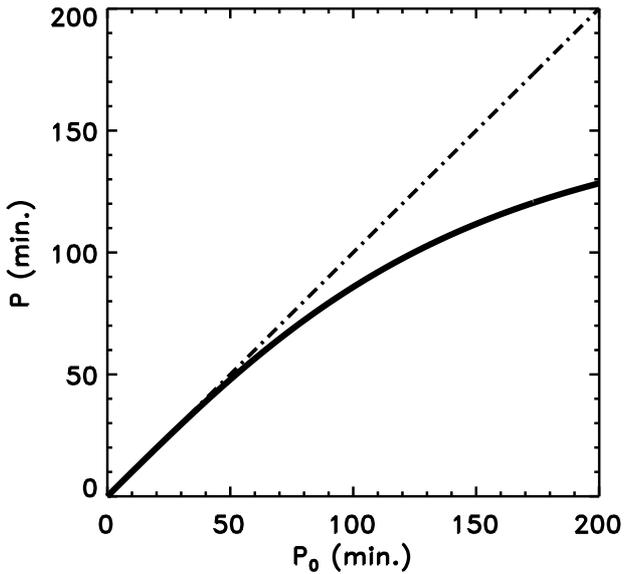}
\caption{Plot of the corrected pendulum period Eq. \eqref{eq:period2} as a function of the old pendulum period $P_{0}$ (solid line). The dot-dashed line is plotted to show the diagonal where $P=P_{0}$. We see that the period deviates from the diagonal for periods larger than 50 minutes.\label{fig:periods-comparison}}
\end{figure}

The gravity vector is $\vec{g}=-\grav \, (\sin \alpha, \cos \alpha)$, and the tangent unitary vector along the field line is $\vec{u}=(\cos \theta, \sin \theta)$. The scalar product of both vectors gives the projection of the gravity along the field line,
\begin{equation}\label{eq:gravity-projected}
g_\parallel = -\grav \, (\sin \alpha \cos \theta + \cos \alpha \sin \theta) = -\grav \sin (\theta + \alpha) \,.
\end{equation}
With Eq. \eqref{eq:relation-alpha-theta-approximated} we obtain an approximated expression for the projected gravity
\begin{equation}\label{eq:approximated-gravity}
 g_\parallel = -\grav \, \left(\frac{1}{R} + \frac{1}{R_\odot}\right) \, s  \,,
\end{equation}
where we have used the identity $\theta=s/R$. This equation shows the dependence of $g_\parallel$ on position $s$. On the right side there are two contributions. The first is associated with the change of magnetic-field direction along the line and therefore the change of projection of gravity. The second is associated with the intrinsic variation of nonuniform gravity. In an ideal situation with a straight horizontal tube, $R=\infty$, the projection of gravity is not zero. This contrasts with the uniform gravity case, $\rsun=\infty$, where the projection is zero.
Introducing Eq. \eqref{eq:approximated-gravity} into \eqref{eq:wave-eq-harmonic} yields
\begin{equation}\label{eq:wave-eq-final}
c_{s}^{2} \frac{\partial^{2} v}{\partial s^{2}} -\gamma \grav s \left(\frac{1}{R}+\frac{1}{R_\odot} \right)\frac{\partial v}{\partial s} + v \left[\omega^{2}-\grav\, \left(\frac{1}{R}+\frac{1}{R_\odot}\right) \right] =0 ~,
\end{equation}
In this modified equation of motion, the curvature of the flux-tube dip and the curvature of the solar surface are included, in contrast to Eq. (7) of Paper I. We define an equivalent radius of curvature as
\begin{equation}\label{eq:equivalent-radius}
\frac{1}{R_\mathrm{eq}} \equiv \frac{1}{R} + \frac{1}{R_{\odot}} \, .
\end{equation}
Substituting Eq. \eqref{eq:equivalent-radius} into Eq. \eqref{eq:wave-eq-harmonic}, we obtain a formally equivalent expression to Eq. (9) from Paper I. Taking advantage of this, we find that the oscillation frequency in a semi-circular dip is described by the equivalent to Eq. (25) in Paper I, but now with $R_\mathrm{eq}$ instead of $R$, namely  
\begin{equation}\label{eq:frequency-approximation}
\omega^{2} \approx \frac{\grav}{R} + \frac{\grav}{\rsun} + \omega^{2}_\mathrm{slow} \, ,
\end{equation}
where the first two terms are associated with the gravity and the last $\omega_\mathrm{slow}$ is the slow-mode angular frequency associated with the gas-pressure gradient. In \S\ref{subsec:maximum-period} we present an approximation for this frequency.
In prominences the slow-mode term is negligible compared with the terms associated with the solar gravity remaining in the pendulum model (see Paper I). The period is given by
\begin{eqnarray}\label{eq:period}
P=2\pi \, \sqrt{\frac{R_\mathrm{eq}}{\grav}} = 2\pi \, \sqrt{\frac{1}{\grav \, \left( \frac{1}{R}+ \frac{1}{\rsun}\right)}} \, .
\end{eqnarray}
We define the period with uniform gravity as $P_0 = 2\pi \, \sqrt{{R}/{\grav}}$. Then
\begin{eqnarray}\label{eq:period2}
P=  \frac{P_0}{\sqrt{ 1+ \frac{R}{R_{\odot}}}} \, .
\end{eqnarray}
Hence the period is not affected by the solar curvature, $P\approx P_{0}$, when $R\ll \rsun$.
Fig. \ref{fig:periods-comparison} shows a plot of both periods, $P$ and $P_0$, as functions of the period of the uniform-gravity case, $P_0$. For $R << R_{\odot}$, both periods are almost identical as expected. However, for increasing values of $P_{0}$ the $P$ curve starts to diverge from the diagonal, becoming particularly evident beyond $50$ minutes. This figure shows that the solar-surface curvature introduces a very small corrections for typical LALO periods of around one hour, but the deviation is significant for longer periods. 

\subsection{The existence of a maximum pendulum period}
\label{subsec:maximum-period}
Prominence threads are supported against gravity by the magnetic field. At the bottom of the dips, the Lorentz force points into the vertical direction (i.e. $\vec{\hat{z}}$ in Cartesian coordinates) such that
\begin{equation}\label{eq:lorentz-force}
\vec{j}\times\vec{B} = -\frac{1}{2 \mu} \frac{\partial B^{2}}{\partial z} \, \vec{\hat{z}} + \frac{B^{2}}{\mu R} \, \vec{\hat{z}} \, ,
\end{equation}
where the first term on the right side is the magnetic pressure gradient and the second term is the magnetic tension \citep[see][]{priest_magnetohydrodynamics_2014}. Although the dips are not necessarily semi-circular and the radius of curvature depends on the position, $R$ is the curvature at the bottom of the dip in Eq. \eqref{eq:lorentz-force}. The tension term points upwards whereas the magnetic pressure gradient points downwards. This implies that $\frac{\partial B^{2}}{\partial z}>0$, i.e., the field strength increases with $z$ as observed in the prominence core where the cool plasma is located \citep[e.g.][]{rust_magnetic_1967,leroy_magnetic_1983}.  If the field is approximately force-free, these two terms are nearly balanced. With the presence of the prominence mass, a small excess of magnetic tension provides support against gravity. This additional magnetic tension is given by a perturbation of the magnetic configuration with respect to the force-free configuration. 

\begin{figure}[!h]
\centering\includegraphics[width=0.5\textwidth]{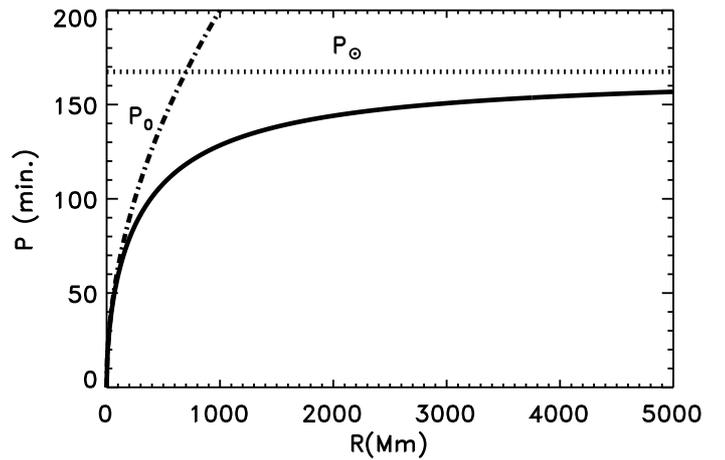}
\caption{Plot of the corrected period (Eq. \eqref{eq:period}) as a function of $R$ (solid line). For comparison, the dashed line shows the uncorrected pendulum period $P_{0}$. The horizontal dotted line shows the cut-off period $\psun$ that $P$ tends toward asymptotically. \label{fig:period-asymptotic}}
\end{figure}

Eq. \eqref{eq:lorentz-force} shows that the support against gravity is given by the curvature of the dipped part of the flux tubes. The only possibility to have support is for positive values of the radius of curvature, $R>0$. For negative values of $R$, the magnetic configuration is a loop. In this sense, to have a prominence supported against gravity $R \in (0,\infty)$ where $R=\infty$ corresponds to a straight line. By substituting progressively increasing values of $R$ into Eq. \eqref{eq:period}, we see that the term $1/R$ tends to zero and the period, $P$, tends asymptotically to $2\pi \, \sqrt{\rsun/\grav}$ and never exceeds this value because when $R$ approaches infinity there is no dip and therefore no possibility of support against gravity. The cut-off frequency corresponds to a period of
\begin{equation}\label{eq:period-sun}
\psun=2\pi \, \sqrt{\frac{\rsun}{\grav}}=167 \mins \, .
\end{equation}
Fig. \ref{fig:period-asymptotic} shows the period as a function of the radius of curvature of the dip using the pendulum approximation \eqref{eq:period}. For large values of $R$ the period curve approaches $\psun$, whereas $P_{0}$ increases monotonically with $R$. 

\subsection{Effect of gas pressure}
\begin{figure}[!h]
\centering\includegraphics[width=0.5\textwidth]{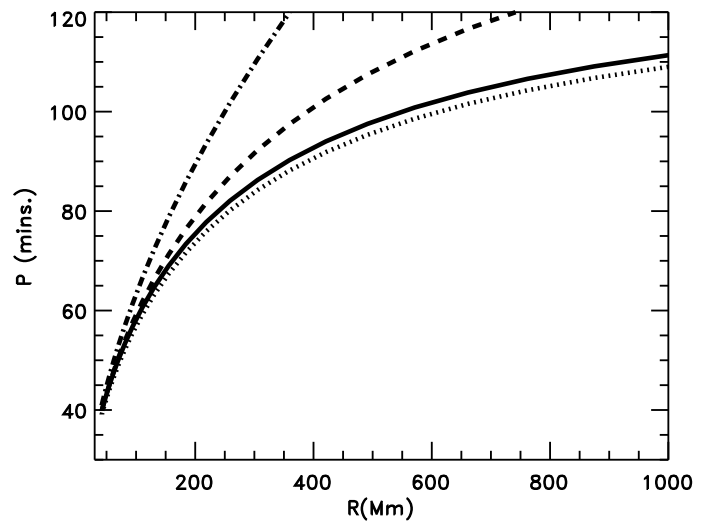}
\caption{Plot of the period of the fundamental-mode period found by solving Eq. \eqref{eq:wave-eq-harmonic} for Model 1 (solid line). The dotted line is the approximation given by Eq. \eqref{eq:period-approximation}, the dashed line shows the corrected pendulum period from Eq. \eqref{eq:period}, and the dot-dashed line is the uncorrected pendulum-model period $P_{0}$.
\label{fig:normalmodes-periods-circular}}
\end{figure}

The previous analysis is valid when the pendulum approximation is applicable. However, in general, the contribution of the gas pressure should be considered. To find the normal modes of flux tube Model 1 (Fig. \ref{fig:fluxtube-geometries}(a)), we solve numerically Eq. \eqref{eq:wave-eq-harmonic}.  Line-tying conditions are imposed at the footpoints, and the eigenvalue problem is solved by means of a shooting technique. 
The numerical routine provides the eigenfunction and the corresponding eigenfrequency of the different modes allowed in the system \citep[see further details in ][]{terradas_magnetohydrodynamic_2013}. 
Fig. \ref{fig:normalmodes-periods-circular} shows the periods of the fundamental normal mode computed numerically, the corrected pendulum period from Eq. \eqref{eq:period}, and the original (uncorrected) pendulum period $P_0$. 

All periods shown in the figure are similar for small $R$ and increase with radius of curvature. For relatively large $R$ values, however, the period $P$ differs considerably from the corrected pendulum approximation because $P$ is significantly modified by the slow-mode contribution in this model flux tube. We have also plotted the period given by the approximation from Eq. \eqref{eq:frequency-approximation}. From this equation the period is
\begin{equation}\label{eq:period-approximation}
\frac{1}{P^{2}}=\frac{\grav}{4\pi^{2}\,R}+\frac{1}{\psun^{2}}+\frac{1}{\pslow^{2}} \, .
\end{equation}
The period $\pslow=2\pi/\omega_\mathrm{slow}$ is computed numerically by solving Eq. \eqref{eq:wave-eq-final} assuming $\grav=0$ and finding the normal mode. In this configuration the slow-mode period is independent of the field-line geometry and equal to $\pslow=206\mins$.
We derive an approximate expression for $\pslow$ by modifying Eq. (26) from Paper I, namely
\begin{equation}
\pslow=\frac{2 \, \pi}{\omega_\mathrm{slow}}\sim 2\,\pi  \sqrt{\frac{\hat{l}\left(L_{1/2}-\hat{l} \right) \, \chi}{c_{sc}^2}}  \, ,  
\end{equation}
where $\chi=T_c/T_p$ is the temperature ratio between the coronal plasma and the prominence. $c_{sc}$ is the sound speed at the corona with temperature $T_c$.
In Paper I the parameter $\hat{l}$ equals the half-length of the prominence $l_{1/2}$ because there is a sharp transition between prominence and corona. Here, in contrast, we have a smooth transition between both media given by Eq. \eqref{eq:temperature}. Using $\hat{l}=l_{1/2}+1/3 \, l_\mathrm{tr}$ approximates very well the exact value of $\pslow$.
The period $P$ containing all contributions agrees very well with the exact numerical solution (dotted line in Fig. \ref{fig:normalmodes-periods-circular}). From Eq. \eqref{eq:period-approximation} we find that the period approaches the following cut-off value as $R\to\infty$:
\begin{equation}\label{eq:cutoff}
	\pcutoff=\frac{\psun \pslow}{\sqrt{\pslow^2+\psun^2}} \, .
\end{equation}
For small values of $\pslow$, $\pcutoff\approx \pslow$. In contrast, for $\pslow\gg\psun$, $\pcutoff\to\psun$, and $P<\pcutoff<\psun$.  In this configuration $\pslow=206\mins$ computed numerically and thus $\pcutoff=130\mins$. 
Eq. \eqref{eq:period-approximation} shows that the longitudinal oscillation period cannot be larger than $\psun$. It is interesting that \citet{jing_periodic_2006} found an event with a period of 160 minutes, which is close to $\psun$. Therefore $\pslow$ should be much larger than the cut-off period in that event. Prominence oscillations with very long (5-6 hours) and ultra-long (up to 30 hours) periods have been reported \citep{foullon_detection_2004,pouget_oscillations_2006,foullon_ultra-long-period_2009}.
The existence of $\pcutoff$ implies that these very low frequency oscillations cannot be attributed to the fundamental magnetoacoustic-gravity mode.

Longitudinal oscillations are governed by two possible restoring forces: the projected gravity and the gas-pressure gradients. In Paper I we found that the gravity dominates when $R\ll R_\mathrm{lim}$ in a situation of uniform gravity, where $R_\mathrm{lim}=\hat{l}\left(L_{1/2}-\hat{l} \right) \, \chi \, \grav /c_{sc}^2$ depends on different parameters of the flux tube and the thread. For the case of nonuniform gravity the relation becomes $R_\mathrm{eq}\ll R_\mathrm{lim}$. Using Eq. \eqref{eq:equivalent-radius} we obtain a new condition on the radius of curvature of the field lines as $R\ll\newrlim$ where
\begin{equation}\label{eq:rlim-condition}
\newrlim = \frac{R_\mathrm{lim}}{|1-\frac{R_\mathrm{lim}}{\rsun}|} \, .
\end{equation}
This is easier to fulfil than the equivalent condition for uniform gravity. With the parameter values stated above, $R_\mathrm{lim}\approx 1100\Mm$, and the new maximum radius is $\newrlim = 1925\Mm$.

Fig. \ref{fig:normalmodes-functions}(a) shows the eigenfunctions, $v=v(s)$, obtained by solving Eq. \eqref{eq:wave-eq-harmonic} numerically. We find a clear dependence on the radius of curvature. For the most curved (smallest $R$) dip in Model 1, the velocity has a wide plateau around the center, extending across the entire dip. In contrast, for an almost flat flux tube, the velocity is mainly concentrated in the dense thread region at the bottom of the dip and the shape of the eigenfunction is almost triangular. Ranging from small to large values of $R$, the eigenfunction continuously changes from a plateau to a triangular shape.
We have considered the same case without the effect of the solar surface curvature, i.e. $\rsun=\infty$. Fig. \ref{fig:normalmodes-functions}(a)
shows the velocity profile for the normal modes of both maximum and minimum values of $R_0$ with uniform gravity. The case with the minimum $R_0$ is very similar to the thick solid line. Similarly, the case with the maximum $R_0$ is almost identical to the thick grey line. This indicates that there are no important differences between the eigenfunctions in both situations. Therefore, although the effect of the solar curvature modifies the oscillation period, it does not significantly affect the velocity profile of the fundamental normal mode. 

\section{Alternative geometries}\label{sec:tube-geometries}

We compute the oscillations of the fundamental mode in Models 2 and 3, in which the central segment assumes semi-elliptical and sinusoidal shapes, respectively. We compare the period predicted by the pendulum model with the period corresponding to the radius of curvature at the center of the dip, $R_0$, given by Eqs. \eqref{eq:radiusofcurvature-ellipse} and \eqref{eq:radiusofcurvature-sinusoidal} for these models.

\begin{figure}[!h]
	\centering\includegraphics[width=0.5\textwidth]{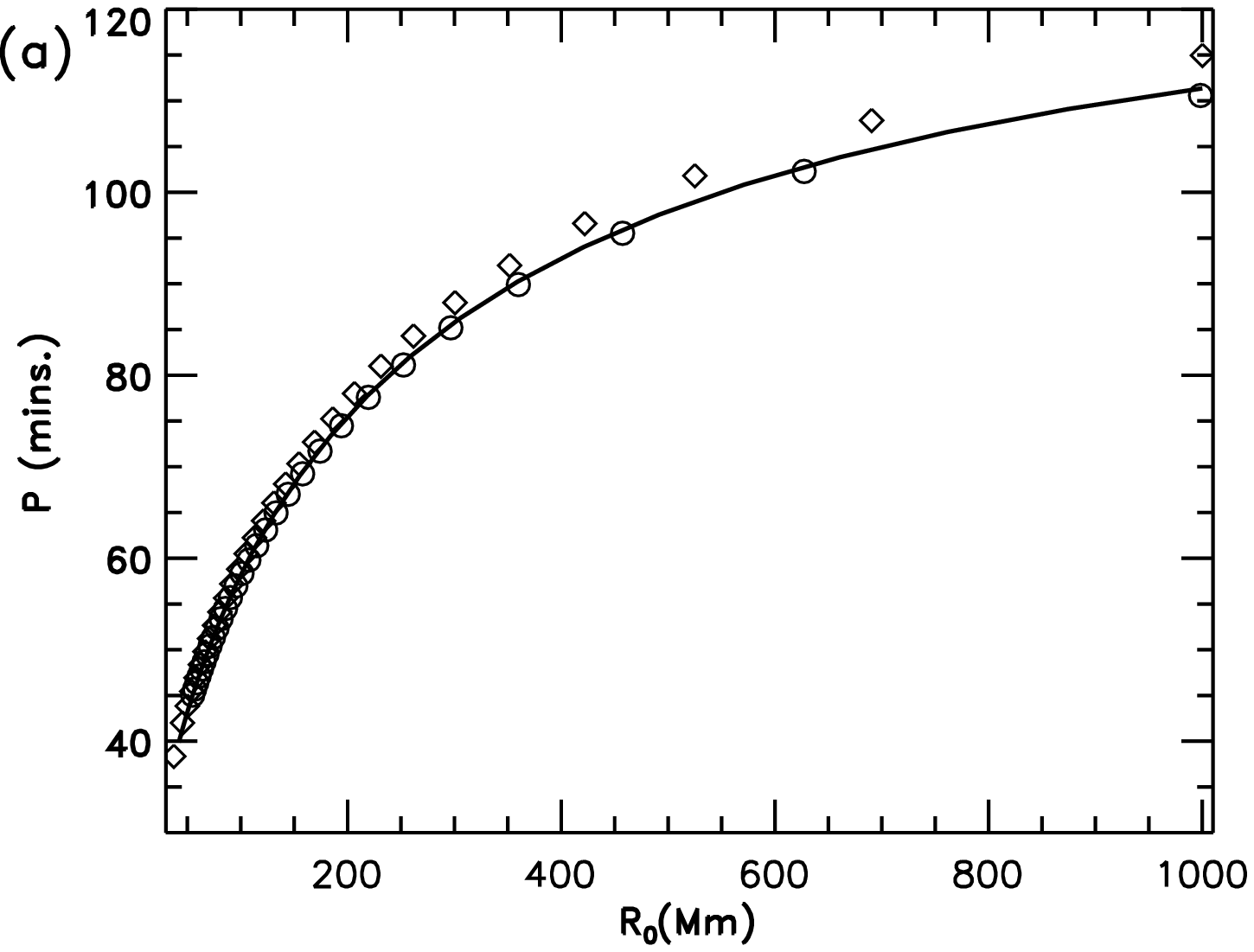}
	\centering\includegraphics[width=0.5\textwidth]{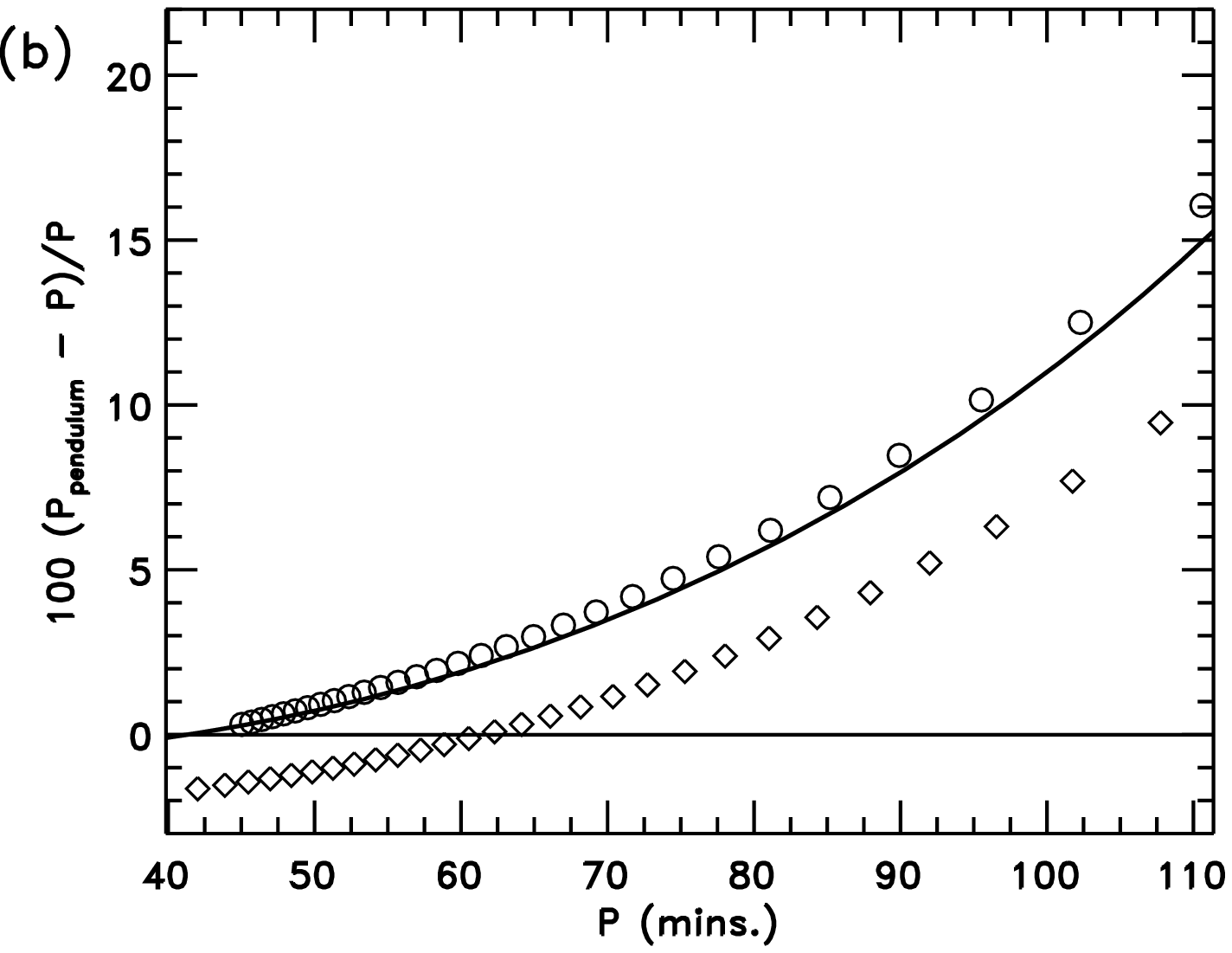}
	\caption{(a) Plot of the period as a function of the radius of curvature at the bottom of the dip, $R_{0}$, for the three geometries considered: Model 1 (solid line), Model 2 (circles), and Model 3 (diamonds). (b) The percent discrepancy between the exact period and that estimated by the corrected pendulum model (Eq. \eqref{eq:period}) as a function of the exact period. \label{fig:normalmodes-periods-comparison}}
\end{figure}

Fig. \ref{fig:normalmodes-periods-comparison}(a) shows the period as a function of the radius of curvature at the bottom of the dip for the three models considered in this work. 
The periods of the semi-circular and semi-elliptical dip geometries are very similar. Model 2 has periods slightly smaller than the semi-circular case, while the Model 3 periods are larger than for Model 1. In both cases, the discrepancies are very small for small radii of curvature and increase with $R_{0}$. This figure shows that the periods are roughly independent of the flux-tube geometry for periods below 120 minutes.

\begin{figure}[!h]
\centering\includegraphics[width=0.5\textwidth]{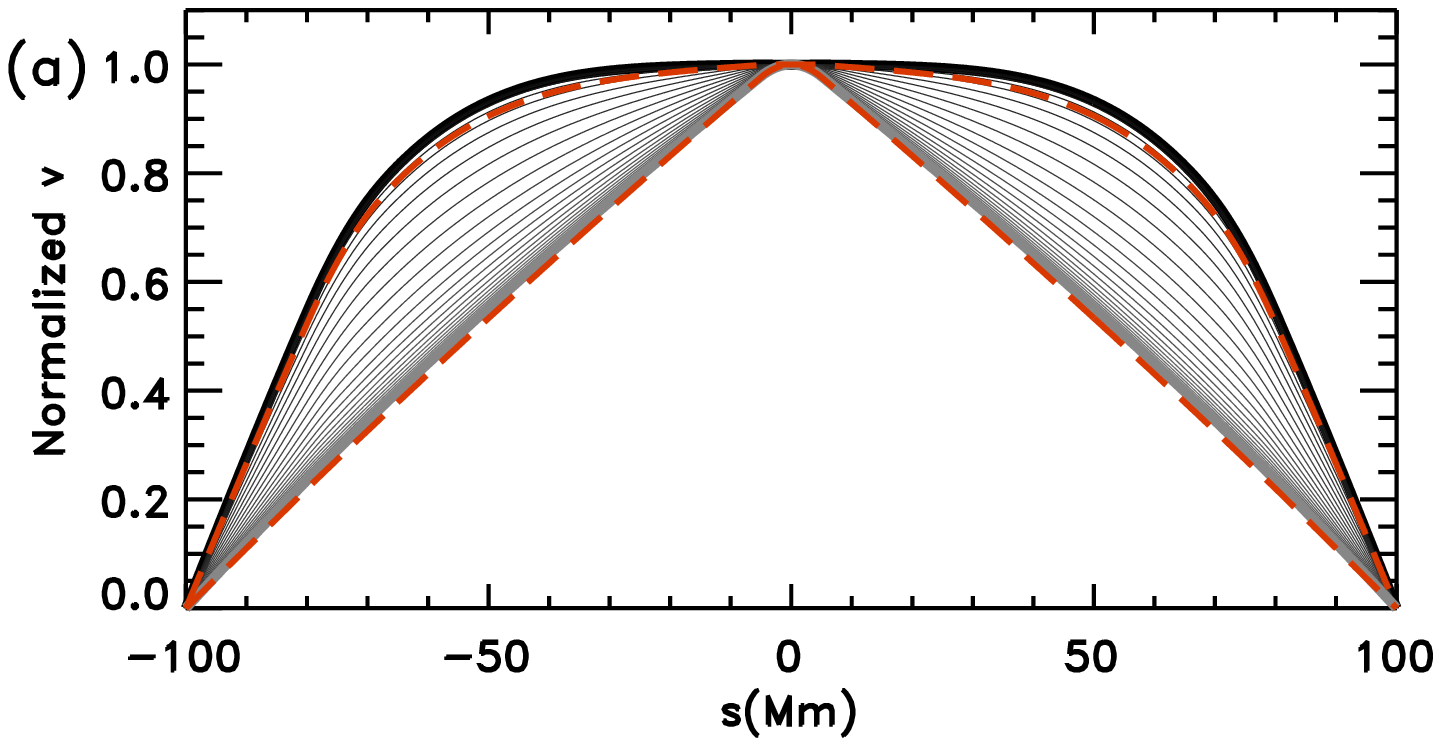}
\centering\includegraphics[width=0.5\textwidth]{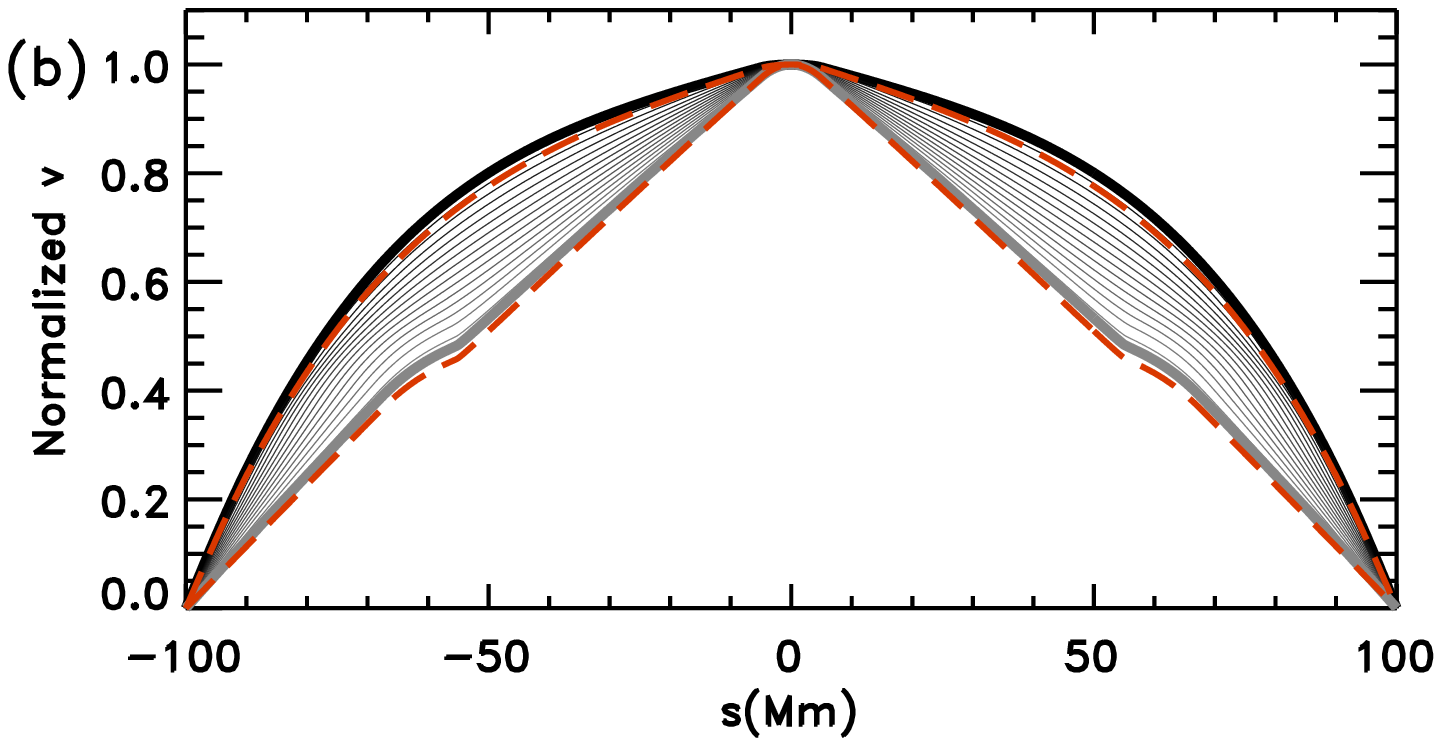}
\centering\includegraphics[width=0.5\textwidth]{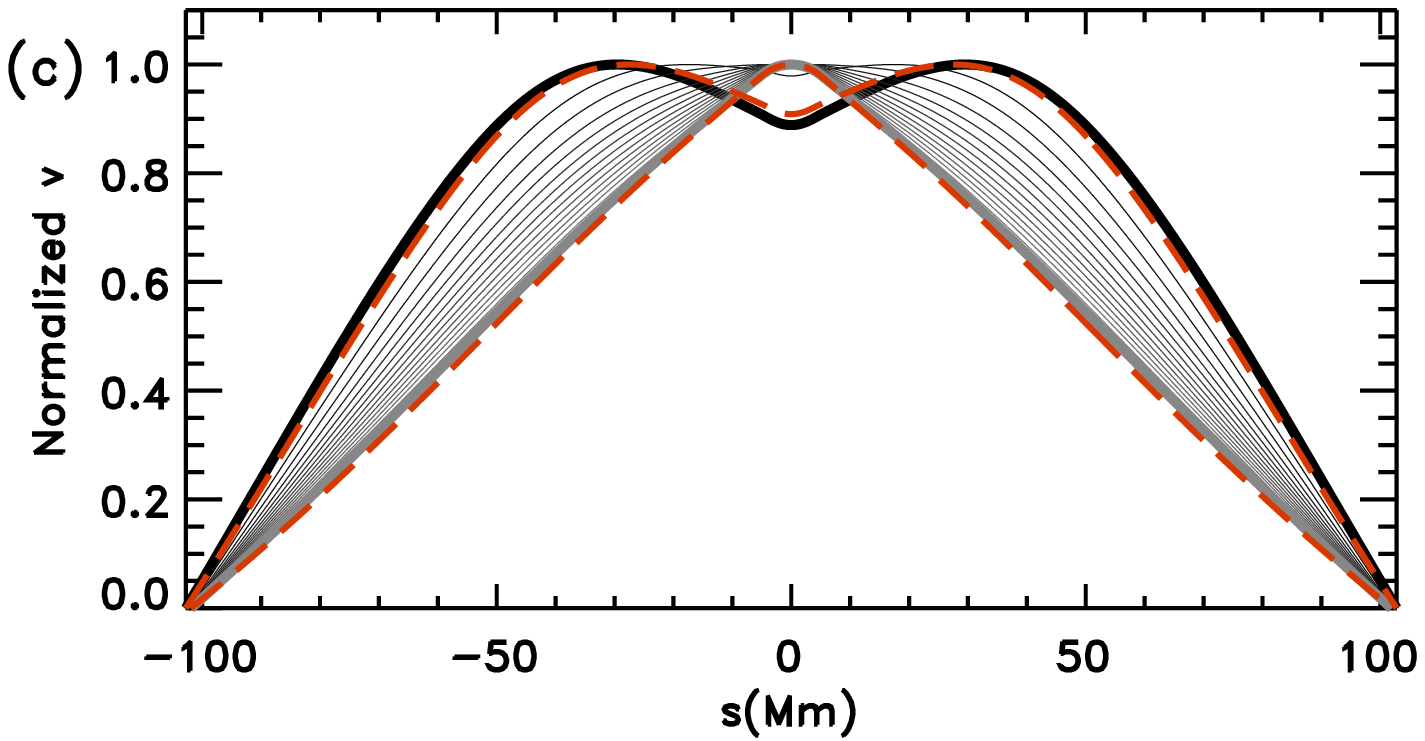}
	\caption{Plot of the eigenfunctions $v=v(s)$ solutions of Eq. \eqref{eq:wave-eq-harmonic} normalised to its maximum value for (a) circular, (b) elliptical, and (c) sinusoidal tube models (see Fig. \ref{fig:fluxtube-geometries}). 
The thick black line corresponds to the mode with the minimum radius of curvature for each model (see \S \ref{subsec:geometry}), and the thick grey line corresponds to $R_{0}=1000\Mm$. 
The thin lines correspond to eigenfunctions with intermediate $R_{0}$ values, smoothly transitioning between both extreme functions. The two red dashed lines in (a) are the normal modes for both maxima and minima of $R_0$ for the uniform-gravity case. 
\label{fig:normalmodes-functions}}
\end{figure}

Fig. \ref{fig:normalmodes-periods-comparison}(b) shows the difference between the actual fundamental-mode period, $P$, and the corrected pendulum-model period $P_\mathrm{pendulum}$ given by Eq. \eqref{eq:period}, as a function of the period $P$. This difference is expressed in \% as $100 (P_\mathrm{pendulum}-P)/P$. The discrepancy with the pendulum model is below 5\% for periods less than 75 minutes, and  increases with the period. For Models 1 and 2, the discrepancies are below 10\% for periods under 95 minutes. In contrast, for the sinusoidal geometry (Model 3), the difference is below 10\% in all of the displayed domain. 
The differences between the curves are not very significant, which leads us to conclude that the range of validity of the pendulum model is similar for all geometries.

The eigenfunctions $v=v(s)$ for Models 2 and 3 are shown in Figs. \ref{fig:normalmodes-functions}(b) and \ref{fig:normalmodes-functions}(c). The functions for the semi-elliptical case are similar to those of the semi-circular geometry (Fig. \ref{fig:normalmodes-functions}(a)), with the maximum velocity located at the tube center. For small $R$ the function is very flat, but there is no clear plateau as in Model 1. For flatter tubes with larger eccentricity, the motion is more and more confined in the cool thread; the eigenfunction shape becomes more triangular as the eccentricity increases. The function has a small deviation at around $|s|=55\Mm$ that is an artifact of the transition between the dip and the adjoining straight horizontal tubes.
The Model 3 eigenfunctions differ substantially from those of the other models. For deep dips, the eigenfunction has two maxima not located at the dip bottom. For shallower dips, the function changes and the two maxima approach one another as $R$ decreases. Finally, for the flattest tubes, the function resembles a triangle as in previous models. As for Model 1 we have also computed the eigenfunctions in the case of uniform gravity, and found negligible differences with the non-uniform gravity case (see red dashed lines).

\section{Influence on prominence seismology}
\label{sec:seismology}
\begin{figure}[!ht]
	\centering\includegraphics[width=0.5\textwidth]{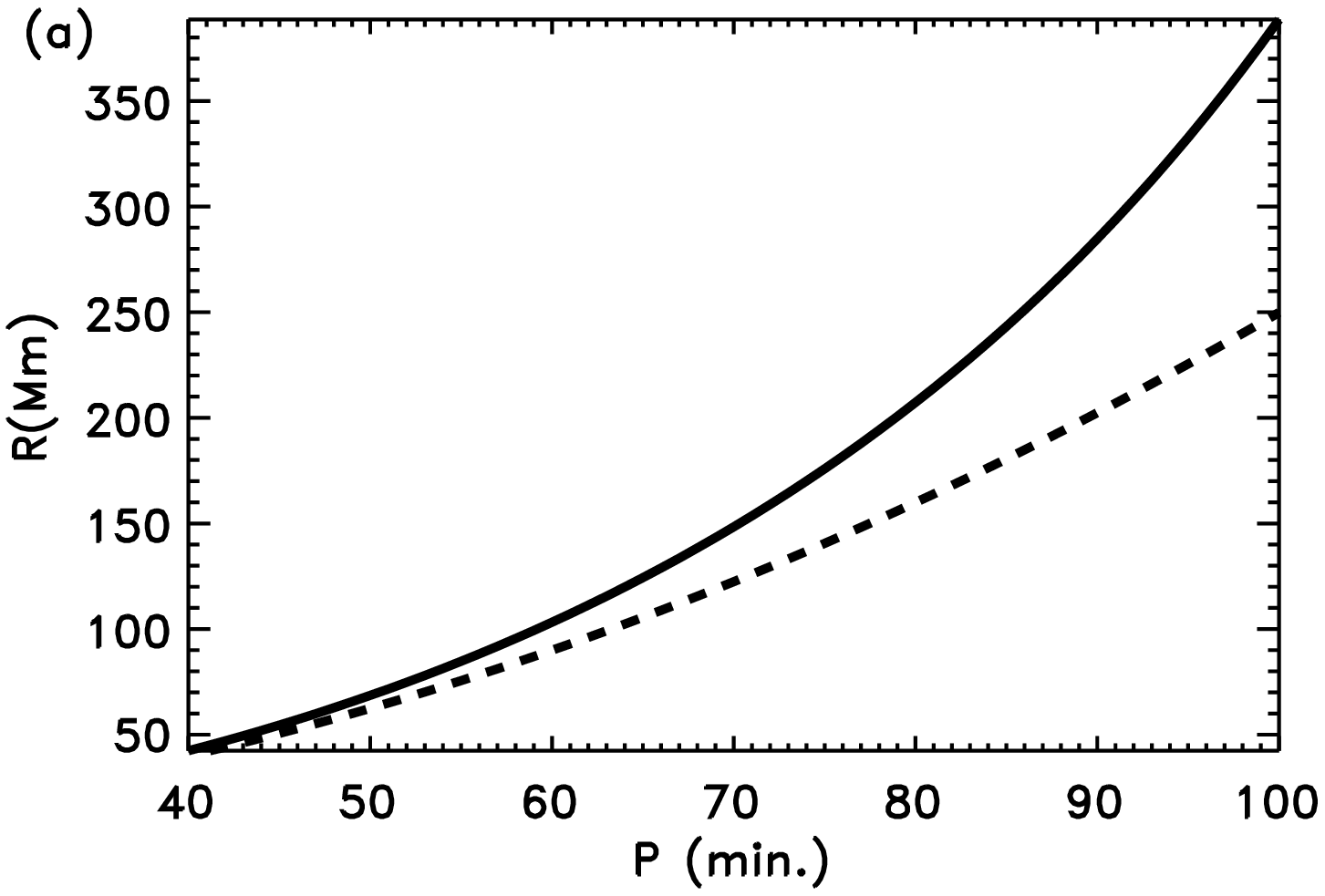}
	\centering\includegraphics[width=0.5\textwidth]{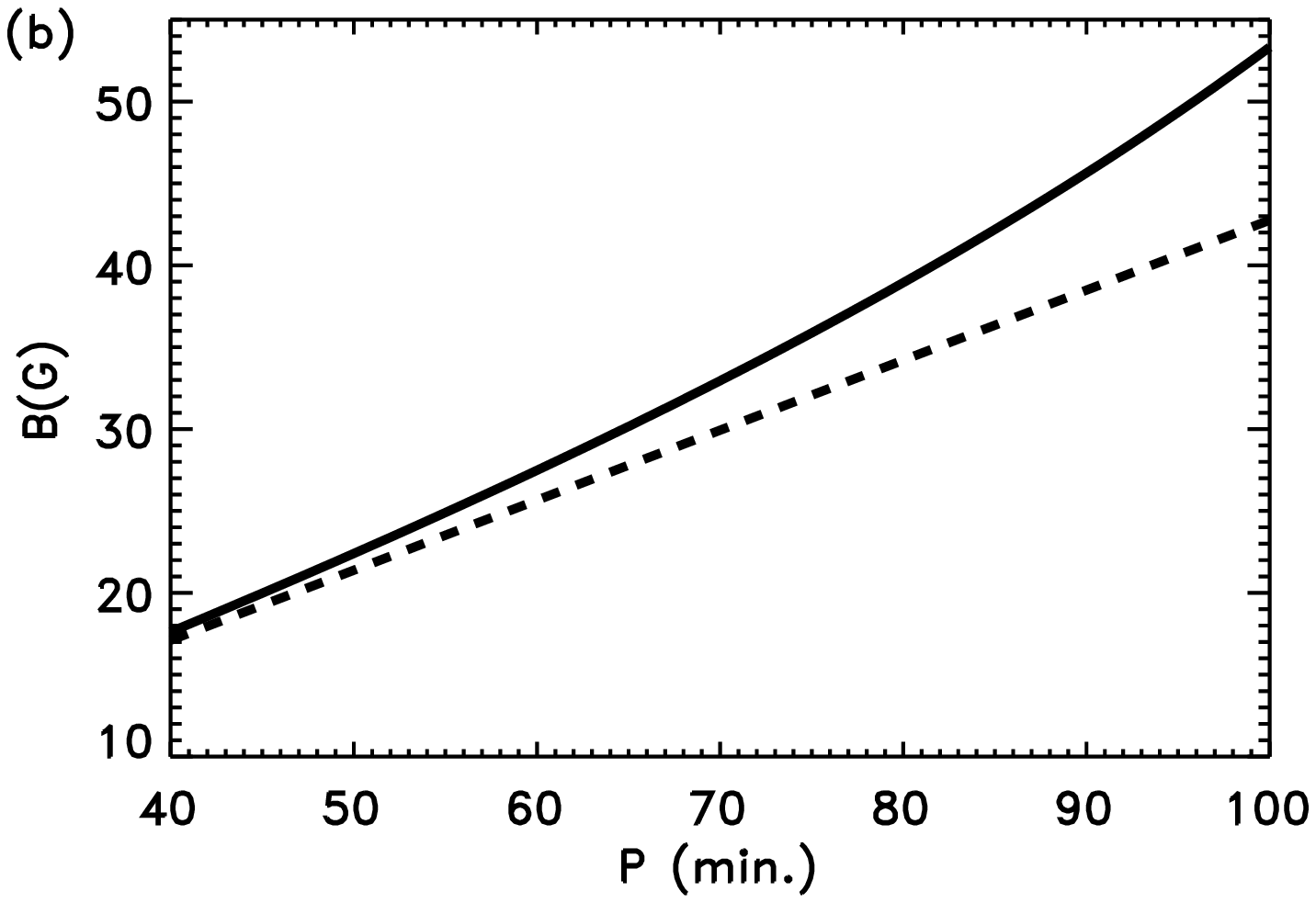}
	\caption{(a) Radius of curvature and (b) the minimum magnetic-field strength vs period, for the old pendulum model (dashed line) and the corrected expression including the curvature of the solar surface (solid line) from Eq. \eqref{eq:seismology}. In this plot we have considered $\rho=2\times10^{-10} ~\mathrm{kg \, m^{-3}}$.\label{fig:seismology}}
\end{figure}

The prominence-seismology technique combines theoretical modelling of prominence oscillations with observations to infer hidden or hard to measure features of the prominence structure.
LALOs provide unique measurements of the curvature of the dips in filament-channel flux tubes, and a minimum value for the magnetic-field strength in those dips \citep[e.g.,][]{luna_large-amplitude_2012}. The relations between the oscillation period and these two parameters have changed from our earlier results, due to the corrections introduced by the intrinsic spatial variation of the solar gravity. Assuming that the pendulum model works well (i.e. $\pslow\to\infty$), Eq. \eqref{eq:period-approximation}  yields a new relation between the radius of curvature and the period:
\begin{equation}\label{eq:new-radius-seismology}
    R=\frac{\grav \, P^2 }{4\pi^2\left[1-\left(\frac{P}{\psun}\right)^2\right]}=\frac{R_\mathrm{old}}{1-\left(\frac{P}{\psun}\right)^2} \, ,
\end{equation}
where $R_\mathrm{old}=\grav P^2/4\pi^2$ is the radius of curvature computed with uniform gravity from Paper I.
Fig. \ref{fig:seismology}(a) shows the new relation between $R$ and $P$. In the same figure we have also plotted $R_\mathrm{old}$ (dashed line). The corrections are increasingly significant for periods longer than 60 minutes. For example, for $P=80$ minutes the radius of curvature is approximately 50 Mm larger than the uncorrected estimate.

The pendulum model also predicts a minimum value for the magnetic field strength of the prominence (Paper I), by assuming that the magnetic field supports the cool, dense prominence threads. As shown in \S\ref{subsec:maximum-period}, the magnetic tension is responsible for the support. Thus, according to Eq. \eqref{eq:lorentz-force}, the magnetic tension is larger than the weight of the threads, such that
\begin{equation}\label{eq:magneticsupport-condition}
\frac{B^{2}}{\mu \, R} -\rho \, g \ge 0 \, .
\end{equation}
Assuming that the slow-mode (gas-pressure gradient) contribution is small, Eq. \eqref{eq:new-radius-seismology} yields the following corrected relation

\begin{equation}\label{eq:seismology}
B \ge \sqrt{\frac{\mu \, \rho \, g^{2}}{4\pi^{2}}} \frac{P}{\sqrt{1-\left( \frac{P}{\psun}\right)^{2}}} = \frac{B_\mathrm{old}}{\sqrt{1-\left( \frac{P}{\psun}\right)^{2}}}\, ,
\end{equation}
where $B_\mathrm{old}$ is defined by Eq. (7) in Paper I. The difference comes from the new term in the denominator. Fig. \ref{fig:seismology}(b) compares the minimum magnetic-field strength inferred from the corrected pendulum model with the values obtained with the uncorrected version. The discrepancy increases with the period; in fact, Eq. \eqref{eq:seismology} shows that the corrected field strength increases asymptotically as $P$ approaches $\psun$. For periods larger than $\psun$ the equation is not valid. However, for typical LALO periods \citep[$\le90$ minutes;][]{luna_gong_2018}, the discrepancies are small. As we noted previously \citep{luna_observations_2014}, the density $\rho$ introduces an important uncertainty in Eq. \eqref{eq:seismology} unless $\rho$ is directly measured, because the range of possible values is two orders of magnitude. Therefore the uncertainty in the magnetic-field determination is much larger than the correction introduced by the non-uniform gravity, for typical LALO periods ($P\sim 1 $ hour). For longer periods, however, the intrinsic spatial variations of the gravity might introduce comparable corrections.

\section{Discussion and Conclusions}\label{sec:conclusions}

In this work, we show that the longitudinal oscillations (magnetoacoustic-gravity modes) commonly observed in solar prominence threads are influenced by the intrinsic spatial variations of the solar gravity, which always points radially toward the solar centre.
This effect introduces a correction in the equations governing the longitudinal oscillations, and modifies the pendulum-model approximation. This correction is significant for periods larger than 60 minutes and increases with increasing period.
A new radius of curvature is defined as the combination of the dip radius of curvature and the solar-surface curvature.
The gravity correction has another interesting effect. In order to support the prominence against gravity, the dips of the field lines must have concave-upward curvature; the limiting case is straight lines, with no curvature. Hence a cut-off period exists for longitudinal oscillations. 
Observed longitudinal oscillations always have periods below $P<\psun=167\mins$, which can be tested observationally. The largest longitudinal period reported so far is 160 minutes \citep{jing_periodic_2006}, very close to the cut-off. This confirms that the main restoring force is gravity, and that the gas-pressure gradients make only small contributions to the oscillations. In addition, prominence oscillations with periods between 5 and 30 hours have been detected \citep{pouget_oscillations_2006,foullon_detection_2004,foullon_ultra-long-period_2009}. The existence of the cut-off period excludes the possibility that those oscillations are fundamental magnetoacoustic-gravity modes.

The pendulum model approximation assumes that the prominence dips are semi-circular, for simplicity. In this work, we have studied the influence of the dip geometry on magnetoacoustic-gravity modes by modelling flux tubes with semi-circular, semi-elliptical, and sinusoidal dips. 
In all the cases the period depends mainly on the radius of curvature at the bottom of the dip and not on the exact model considered. 
We found that the pendulum model is quite robust and is still valid for non-circular dips.

We also studied the influence of the new modified pendulum model on prominence seismology. The new model relates the oscillation period and the minimum magnetic-field strength to the radius of curvature at the bottom of the dips. We found that the magnetic-field correction is not very large for typical observed longitudinal oscillation periods around one hour. However, for periods above 60 minutes the correction is significant. 
\citet{zhang_simultaneous_2017} reported a LALO event with a relatively long period of $\approx$99 minutes, yielding a radius of curvature of 244 Mm and a field intensity of 28 G under the original pendulum model. With the corrections discussed in this paper, the estimated radius of curvature is 376 Mm and the field intensity is 35 G, which differ substantially from the Zhang et al. values.
More recently, \cite{dai_oscillations_2021} also reported a LALO event with an even longer period of $\approx$120 minutes, from which they estimated a radius of curvature of 355 Mm and a field strength of 34 G with the old pendulum expression. With the corrected pendulum model the radius of curvature is 734 Mm and the field intensity is 49 G.
We conclude that the intrinsic spatial variations of the solar gravity introduce important corrections to the pendulum model. In addition, the corrected pendulum model provides a good estimate of the radius of curvature at the bottom of the dips for any flux-tube geometry.
This work has been limited to modelling linear motions in the oscillations, for which the center of mass of the cool prominence thread moves negligibly around the bottom of the dip. However, most longitudinal oscillations are classified as large amplitude, where the prominence thread is advected far from the dip center through regions with different curvatures. In order to understand in-depth the relationship between the period and the flux-tube geometry, this nonlinear behaviour  will be the subject of future research.

\begin{acknowledgements}
M.L. acknowledges support through the Ram\'on y Cajal
fellowship RYC2018-026129-I from the Spanish Ministry of
Science and Innovation, the Spanish National Research Agency
(Agencia Estatal de Investigaci\'on), the European Social Fund
through Operational Program FSE 2014 of Employment,
Education and Training and the Universitat de les Illes Balears. This publication is part of the R+D+i project PID2020-112791GB-I00, financed by MCIN/AEI/10.13039/501100011033.
J.T.K. was supported in part by NASA's H-ISFM program. The authors also acknowledge support from the International Space Sciences Institute (ISSI) via team 413 on ``Large-Amplitude
Oscillations as a Probe of Quiescent and Erupting Solar
Prominences''
\end{acknowledgements}

%\bibliography{curvature.bib}

\end{document}